\documentclass[article, twocolumn,
   aps, pra,
  amsmath,amssymb,
  longbibliography,
  ]{revtex4-1}
\usepackage{graphicx,color}
\usepackage{amsmath}
\usepackage{natbib}
\usepackage{epsfig}
\begin{document}

\title{\color{blue} Minima of shear viscosity and thermal conductivity coefficients of classical fluids}

\author{S. A. Khrapak}\email{Sergey.Khrapak@gmx.de}
\affiliation{Joint Institute for High Temperatures, Russian Academy of Sciences, 125412 Moscow, Russia}
\author{A. G. Khrapak}
\affiliation{Joint Institute for High Temperatures, Russian Academy of Sciences, 125412 Moscow, Russia}

\begin{abstract}
{The shear viscosity and thermal conductivity coefficients of various liquids exhibit minima along certain trajectories on the phase diagram. These minima arise due to the crossover between the momentum and energy transport mechanisms in gas-like and liquid-like regimes.} We demonstrate that the magnitudes of the minima  are quasi-universal in appropriately reduced units, especially for the viscosity coefficients. Results presented in support of this observation concern the transport properties of three simple model systems with different pairwise interaction potentials (hard spheres, Lennard-Jones, and Coulomb) as well as seven important real atomic and molecular liquids (Ne, Ar, Kr, Xe, CH$_4$, CO$_2$, and N$_2$). The minima in viscosity and thermal conductivity represent useful reference points for fluid transport properties. 
\end{abstract}

\date{\today}

\maketitle

\section{Introduction}

It is well recognized that the viscosity and thermal conductivity of liquids are both strongly system dependent, can vary across many orders of magnitude, and cannot be evaluated from first-principles theories~\cite{TrachenkoSciAdv2020}. Recently, it has been suggested that the {{\it kinematic viscosity} and {\it thermal diffusivity}} of liquids and supercritical fluids have lower bounds determined by fundamental physical constants~\cite{TrachenkoSciAdv2020,TrachenkoPRB2021},
\begin{equation}\label{bonds}
\nu_{\rm min}=\alpha_{\rm min}=\frac{1}{4\pi}\frac{\hbar}{\sqrt{m_e m}}, 
\end{equation}
where $\nu$ is the kinematic viscosity, $\alpha$ is the thermal diffusivity, $\hbar$ is the Planck's constant, $m_e$ is the electron mass and $m$ is the atom or molecule mass.
The very existence of universal bounds and their closeness is a notable result, since our understanding of transport properties in fluids remains too far from complete as compared to situations with gases and solids, although some definite progress has been achieved over the years~\cite{FrenkelBook,HansenBook,GrootBook,MarchBook,BalucaniBook}.

The purpose of this paper is to concentrate on purely classical arguments, which suggest that properly reduced viscosity and thermal conductivity coefficients can be expected to reach quasi-universal values at their respective minima. Minima appear because of the crossover between the gas-like and liquid-like mechanisms of momentum and energy transfer. Their magnitudes can be estimated by extrapolating the gas-like and liquid-like asymptotes for the transport coefficients into the crossover regime. This procedure is supported by evidence from several model and real atomic and molecular systems. The observed quasi-universality is particularly striking for the viscosity coefficient. We deal exclusively with  classical fluids. Fluids with considerable electron contribution to the thermal conductivity (e.g. liquid metals or multicomponent plasmas) are not considered in this study.

\section{Normalization} 

Variations of the viscosity and thermal conductivity coefficients with temperature of noble, molecular, and network
liquids along several selected isobars are shown in Fig.~1 of Ref.~\cite{TrachenkoSciAdv2020} and Fig.~1 of Ref.~\cite{TrachenkoPRB2021}. Minima are clearly observed. However, the numerical values of transport coefficients for different substances and different conditions can differ by orders of magnitude. To harmonize the picture, it makes sense to use a rational normalization. A particularly useful approach is to employ a system-independent normalization extensively used by Rosenfeld~\cite{RosenfeldPRA1977,RosenfeldJPCM1999} (and therefore often referred to as Rosenfeld's normalization), 
\begin{equation}\label{Rosenfeld}
D_{\rm R}  =  D\frac{n^{1/3}}{v_{\rm T}} , \quad\quad
\eta_{\rm R}  =  \eta \frac{n^{-2/3}}{m v_{\rm T}}, \quad\quad \lambda_{\rm R}=\lambda\frac{n^{-2/3}}{v_{\rm T}},
\end{equation}
where $D$, $\eta$, and $\lambda$ are the self-diffusion, shear viscosity, and thermal conductivity coefficients and the subscript ${\rm R}$ denotes Rosenfeld's normalization. {Here $n$ is the atomic or molecular density, $v_{\rm T}=\sqrt{k_{\rm B}T/m}$ is the thermal velocity, $T$ is temperature, and $k_{\rm B}$ is the Boltzmann constant.} 

\section{Theoretical minima of viscosity and thermal conductivity}

The origin of the minima in the reduced viscosity and thermal conductivity coefficients is the crossover between different mechanisms of the momentum and energy transfer. 

In dilute gases the transport properties are determined by collisions between the constituent atoms. Atoms move along straight trajectories between collisions. The properties of collisions are governed by the mechanism of the interaction between the atoms. In this way the transport coefficients can be evaluated using the Chapman-Enskog theory~\cite{ChapmanBook}, where they are expressed via the transport integrals (momentum and energy transfer cross sections integrated with the Maxwellian velocity distribution function). Elementary kinetic formulas for the viscosity and thermal conductivity coefficients of dilute gases are~\cite{LifshitzKinetics}
\begin{equation}\label{gas}
\eta \sim mv_{\rm T} n\ell, \quad \lambda\sim (c_p/k_{\rm B})v_{\rm T}n\ell,
\end{equation}   
where $\ell$ is the mean free path between collisions and $c_p$ is the specific heat at constant pressure (numerical coefficients of order unity are omitted). {The mean free path can be expressed via the effective momentum transfer cross section $\Sigma$ as $\ell\sim 1/n\Sigma$ (here we use $\Sigma$ instead of conventional $\sigma$ to avoid confusion with the hard-sphere diameter and Lennard-Jones length scale that will be employed later)}. Moreover, for monatomic dilute gases the exact relation between the viscosity and thermal conductivity exists, $\eta = 4m\lambda/15$, which does not depend on exact mechanisms of interatomic interactions~\cite{LifshitzKinetics}.

In the opposite regime of dense fluids no first principle theoretical formulas are available. However, some approximate relationships do exist. A useful scaling relationship for transport coefficients of fluids is excess entropy scaling proposed by Rosenfeld~\cite{RosenfeldPRA1977}. According to this scaling the reduced viscosity and thermal conductivity coefficients of simple fluids can be expressed as exponential functions of the reduced excess entropy~\cite{RosenfeldJPCM1999}
\begin{equation}\label{Rosenfeld1}
\eta_{\rm R}  \simeq  0.2{\rm e}^{-0.8 (s_{\rm ex}/k_{\rm B})}, \quad \lambda_{\rm R} \simeq 1.5 {\rm e}^{-0.5 (s_{\rm ex}/k_{\rm B})}.
\end{equation}
Here the reduced excess entropy is $s_{\rm ex}=s-s_{\rm id}$, where $s$ is the entropy per particle and $s_{\rm id}$ is the entropy (per particle) of an ideal gas at the same temperature and density. Note that the excess entropy $s_{\rm ex}$ is negative because interactions enhance the structural order compared to that in an ideal gas. This implies that the reduced viscosity and thermal conductivity coefficients increase when approaching to the freezing point. Excess entropy scaling in the form of Eq.~(\ref{Rosenfeld1}) also suggests that the coefficients $\eta_{\rm R}$ and $\lambda_{\rm R}$ can be interrelated. This possibility has been recently examined in Ref.~\cite{KhrapakJETPLett2021}.    

{Rosenfeld's excess entropy scaling is a useful heuristic approach for many simple (and sometimes not so simple) systems, but counterexamples where it does not apply also exist~\cite{KrekelbergPRE03_2009,KrekelbergPRE12_2009,
FominPRE2010}. For a recent review of this topic see e.g. Ref.~\cite{DyreJCP2018}.} 

According to a vibrational model of thermal conductivity in dense fluids~\cite{KhrapakPRE01_2021,Horrocks1960,AllenPRB1994,KhrapakPoP08_2021} the thermal conductivity coefficient can be estimated from
\begin{equation}\label{vibrational}
\lambda\sim (c_{\rm p}/k_{\rm B}) \nu_{\rm E} n^{1/3},
\end{equation}   
where $\nu_{\rm E}$ is the characteristic frequency of atomic vibrations (e.g. Einstein frequency). Dense fluids close to the freezing point can be considered as essentially incompressible. 
In this cases the difference between $c_{\rm p}$ and $c_{\rm v}$ is insignificant and it is more practical to use $c_{\rm v}$ for approximate estimates~\cite{KhrapakPRE01_2021,KhrapakPoP08_2021}.

Let us now use these approximations to estimate the magnitudes of $\eta_{\rm R}$ and $\lambda_{\rm R}$ at their respective minima. Some of the ideas can be traced back to Ref.~\cite{TrachenkoPRB2021}. As already pointed out, the minima of $\eta_{\rm R}$ and $\lambda_{\rm R}$ correspond to the crossover between the gas-like and liquid-like mechanisms of the momentum and energy transfer. First, let us take the gaseous asymptotes (\ref{gas}) and extrapolate them to the point where gaseous approach breaks down. At the micro scale, gas properties are mostly determined by pairwise collisions between atoms or molecules whereas liquid properties are mainly
controlled by collective effects. A natural transition condition between these two regimes can be defined as the point where the effective momentum transfer cross section becomes comparable to the interatomic separation squared, $\Sigma\sim n^{-2/3}$. This condition was in fact literally employed to discriminate between the ``ideal'' (gas-like) and ``nonideal'' (fluid-like) regions on the phase diagram of complex plasmas (using a screened Coulomb potential model)~\cite{KhrapakPRE2004_MT}. Substituting this condition in Eqs.~(\ref{gas}) we readily obtain estimates $\eta_{\rm R}\sim 1$ and $\lambda_{\rm R}\sim (c_{\rm p}/k_{\rm B})$ at their minima. Furthemore, the specific heat at the crossover may be approximated as $c_{\rm v}/k_{\rm B}\sim 2$ {for monatomic fluids~\cite{TrachenkoPRB2021,BrazhkinPRE2012,BrazhkinPRL2013,ProctorPoF2020} (this argument does not apply to hard sphere fluids, where $c_{\rm v}/k_{\rm B}\equiv 3/2$). By either neglecting the difference between $c_{\rm v}$ and $c_{\rm p}$ or applying the ideal gas relation $c_{\rm p}=c_{\rm v}+k_{\rm B}$ we arrive at the  condition $\lambda_{\rm R}\sim 2 - 3$ at the minimum. For molecular liquids the actual values of $c_{\rm v}$ and $c_{\rm p}$ should be higher due to the presence of additional degrees of freedom. As a result the numerical values of $\lambda_{\rm R}$ at the minima should also be larger. Some representative values for molecular liquids will be provided below.}   

Alternatively, we can extrapolate the excess entropy scaling of $\eta_{\rm R}$ and $\lambda_{\rm R}$ in the dense fluid regime towards the crossover point. The crossover corresponds roughly to $s_{\rm ex}/k_{\rm B}\simeq -1$~\cite{KhrapakPRE10_2021}. {For instance, for hard sphere, Lennard-Jones, and inverse-power-law fluids, the minima in reduced shear viscosity occur at an excess entropy approximately equal to $-2k_{\rm B}/3$~\cite{BellJCP2020}; Minima of the kinematic viscosity and crossing of kinetic and potential contributions to viscosity occur at $s_{\rm ex}/k_{\rm B}\simeq -0.9$~\cite{BellJPCL2021}}.  Substituting $s_{\rm ex}/k_{\rm B}= -1$ in Eqs.~(\ref{Rosenfeld1}) we obtain $\eta_{\rm R}\simeq 0.4$ and $\lambda_{\rm R}\simeq 2.5$. This is not too inconsistent with the estimates based on extrapolating the gaseous asymptotes.

Consider now the vibrational model of heat conduction in dense liquids, Eq.~(\ref{vibrational}). The crossover between the gas-like and liquid-like regimes of energy transfer implies $\ell\simeq 1/n\Sigma\simeq n^{-1/3}$, as discussed previously. On the other hand, the oscillation frequency should become comparable with collisional frequency at this point, $\nu_{\rm E}\sim v_{\rm T}/\ell\sim v_{\rm T}n\Sigma\sim v_{\rm T}n^{1/3}$. This implies that the estimates based on gaseous and liquid-like asymptotes coincide at the crossover and predict $\lambda\sim (c_{\rm p}/k_{\rm B})v_{\rm T}n^{2/3}$. In reduced units this amounts to $\lambda_{\rm R}\sim c_{\rm p}/k_{\rm B}$, as already obtained.

In addition, the vibrational model allows for a simple estimate of the thermal conductivity coefficient at the onset of fluid-solid phase transition. According to the celebrated Lindemann’s melting criterion~\cite{Lindemann},  
melting of a three-dimensional (3D) solid occurs when the
atomic vibrational amplitude reaches a threshold value, roughly $\sim 0.1$ of the mean interatomic separation {(Lindemann's criterion is a useful empirical rule, which works relatively well for simple systems, but may fail for more complex interatomic interactions)}. From the energy equipartition we get in the simplest approximation (see e.g.~\cite{KhrapakPRR2020}) 
\begin{equation}
\frac{1}{2}m(2\pi \nu_{\rm E})^2\delta^2 = \frac{3}{2}k_{\rm B}T_{\rm m},
\end{equation}  
where $\delta$ is the vibrational amplitude, $T_{\rm m}$ is the melting temperature. We then assume that $\nu_{\rm E}$ does not vary much upon the phase change and that $\delta\sim 0.1 n^{-1/3}$. This provides an estimate for the Einstein frequency,
\begin{equation}
\nu_{\rm E}\sim \frac{1}{2\pi}\sqrt{\frac{3k_{\rm B}T_{\rm m}}{m \delta^2}}\simeq 2.8n^{1/3}\sqrt{\frac{k_{\rm B}T_{\rm m}}{m}}. 
\end{equation}  
At the freezing point we can estimate specific heats of monatomic fluids as $c_{\rm p}\sim c_{\rm v}\sim 3k_{\rm B}$~\cite{KryuchkovPRL2020}. Substituting all this back into Eq.~(\ref{vibrational}) we obtain that the reduced thermal conductivity coefficient of atomic fluids at the melting temperature can be estimated as
\begin{equation}
\lambda_{\rm R}\sim 2.8 \left(c_{\rm v}/k_{\rm B}\right)\simeq 8.4.
\end{equation}    
Note that the excess entropy scaling of transport coefficients, Eq.~(\ref{Rosenfeld1}), predicts $\eta_{\rm R}\sim 4.9$ and $\lambda_{\rm R}\sim 11.1$ if we assume $s_{\rm ex}/k_{\rm B}\sim -4$ at freezing of simple soft fluids~\cite{RosenfeldPRE2000}.

Thus, from the theoretical perspective we should expect the minima of reduced viscosity and thermal conductivity coefficients at the level of $\eta_{\rm R}\sim 1$ and $\lambda_{\rm R}\sim 3$ (for monatomic fluids). Additionally, the reduced transport coefficients should increase up to $\eta_{\rm R}\sim 5$ and $\lambda_{\rm R}\sim 10$ at the freezing point. We will verify the accuracy of this simple picture below.

\section{Model systems}

The three model systems considered here include hard spheres, Lennard-Jones supercritical fluid and Coulomb (one-component plasma) fluid. The pairwise interaction potentials are quite different in these systems. 

\begin{figure*}
\includegraphics[width=15cm]{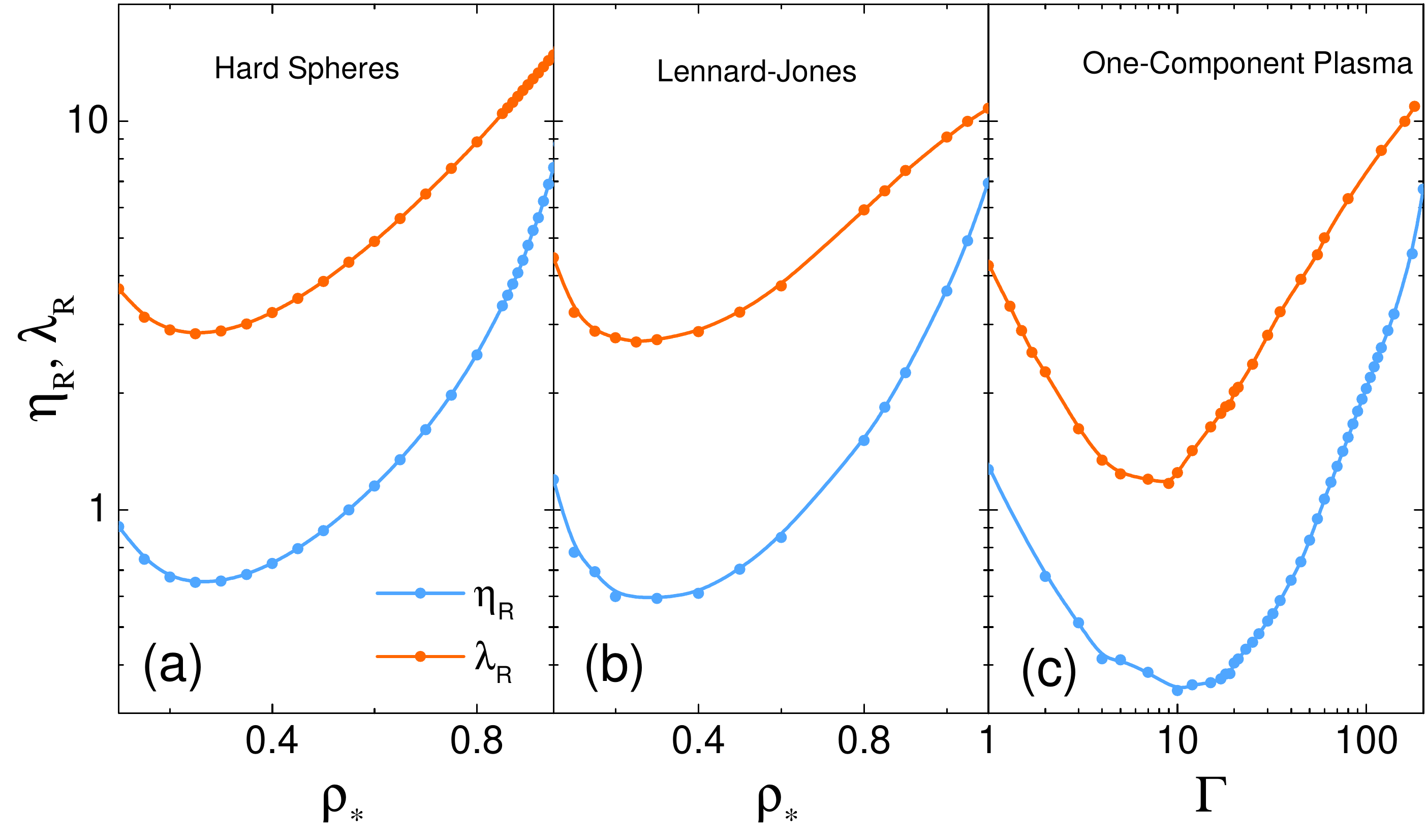}
\caption{(Color online) The reduced shear viscosity $\eta_{\rm R}$ and thermal conductivity $\lambda_{\rm R}$ coefficients of hard sphere (a), Lenard-Jones (b), and Coulomb (c) fluids. The transport coefficients are plotted versus the reduced density $\rho_{*}=\rho\sigma^3$ in (a) and (b) and versus the coupling parameter $\Gamma$ in (c). The symbols connected by smooth curves correspond to the available data: Refs.~\cite{Pieprzyk2019,Pieprzyk2020} for a HS fluid in (a); Refs.~\cite{BaidakovJCP2012,BaidakovJCP2014} for a LJ fluid along an isotherm $T_*=2$ in (b); and Refs.~\cite{DaligaultPRE2014,ScheinerPRE2019} for a Coulomb (OCP) fluid in (c).}
\label{Fig1}
\end{figure*}

The hard sphere (HS) repulsive interaction potential is extremely hard and short ranged. The interaction energy is infinite for $r<\sigma$ and is zero otherwise, where $\sigma$ is the sphere diameter. The main parameter that determines the HS structural and dynamical properties is the reduced density $\rho_*=\rho\sigma^3$. 

The Lennard-Jones (LJ) potential is 
\begin{equation}
\phi(r)=4\epsilon\left[\left(\frac{\sigma}{r}\right)^{12}-\left(\frac{\sigma}{r}\right)^{6}\right], 
\end{equation}
where  $\epsilon$ and $\sigma$ are the energy and length scales (or LJ units), respectively. The reduced density and temperature expressed in LJ units are $\rho_*=\rho\sigma^3$, $T_*=k_{\rm B}T/\epsilon$. The LJ system is one of the most popular and extensively studied model systems in condensed matter, because it combines relative simplicity with adequate approximation of interatomic interactions in real substances (such as liquified and solidified noble gases). {A very comprehensive collection of  data regarding the transport properties of the Lennard-Jones fluid can be founf in Ref.~\cite{BellJPCB2019}.} 
  
The classical one-component plasma (OCP) model is an idealized
system of point charges immersed in a neutralizing 
uniform background of opposite charge (e.g. ions in the immobile background of electrons or vice versa)~\cite{BrushJCP1966,deWitt1978,BausPR1980,IchimaruRMP1982,
KhrapakCPP2016}. This model is of considerable practical interest in both plasma-related and condensed matter context. The system is characterized by a very soft and long-ranged Coulomb interaction potential, 
\begin{equation}
\phi(r)= e^2/r, 
\end{equation}
where $e$ is the electric charge. System properties are governed by a single dimensionless coupling parameter $\Gamma=(e^2/ak_{\rm B}T)$, where $a=(4\pi n/3)^{-1/3}$ is the Wigner-Seitz radius. At $\Gamma\gtrsim 1$, the OCP exhibits properties characteristic of a fluid-like phase and freezes at $\Gamma\simeq 174$~\cite{IchimaruRMP1982,DubinRMP1999,KhrapakCPP2016}. Transport properties of strongly coupled OCP fluids within the vibrational model paradigm have been recently discussed~\cite{KhrapakPRE01_2021,KhrapakMolecules12_2021}.     
 
We summarize the available numerical data for the viscosity and thermal conductivity coefficients of the considered model systems in Fig.~\ref{Fig1}. The transport data for the HS system are taken from recent MD simulations reported in Refs.~\cite{Pieprzyk2019,Pieprzyk2020}. For the LJ system we take the viscosity and thermal conductivity coefficients tabulated in Refs.~\cite{BaidakovJCP2012,BaidakovJCP2014} for an isotherm $T_*=2$ .   
Since the reduced transport coefficients of LJ liquids along isotherms exhibit a quasi-universal freezing-density scaling~\cite{KhrapakPRE04_2021}, the data are also representative to other isotherms. Detailed analysis of LJ fluids transport properties data existing in the literature (in the context of the entropy scaling) can be found in Ref.~\cite{BellJPCB2019}.  For a OCP fluid we use the data from MD simulations tabulated in Refs.~\cite{DaligaultPRE2014,ScheinerPRE2019}.

Figure~\ref{Fig1} demonstrates that the minima of $\eta_{\rm R}$ and $\lambda_{\rm R}$ are clearly observed (it should be noted that the exact location of the minima can depend on the normalization chosen). The magnitudes of transport coefficients are relatively close for HS and LJ fluids: $\eta_{\rm R}\simeq 0.7$ and $\lambda_{\rm R}\simeq 3.0$ at the minimum in HS; $\eta_{\rm R}\simeq 0.6$ and $\lambda_{\rm R}\simeq 2.7$ at the minimum in LJ. For the OCP fluid the minima are significantly deeper: $\eta_{\rm R}\simeq 0.3$ and $\lambda_{\rm R}\simeq 1.2$. We believe that this difference is due to extremely soft and long-ranged character of the Coulomb potential. 

{In fact, the effect of the potential softness on the location and magnitude of the minima in the reduced shear viscosity coefficient was documented in Ref.~\cite{BellJCP2020}. In this work the viscosity coefficient was evaluated in the vicinity of the minima for the inverse-power-law (IPL) repulsive potential family $\phi(r)=\epsilon(\sigma/r)^n$, with various IPL exponent $n= 6,9,12,15,18,24,36,48,52$. It was observed that the location of the minima in terms of excess reduced entropy change almost monotonously from $s_{\rm ex}/k_{\rm B}\simeq -0.71$ for $n=6$ to the hard sphere limiting value of $s_{\rm ex}/k_{\rm B}\simeq -2/3$ for $n=52$. The magnitude of the reduced shear viscosity coefficient at its minimum increases smoothly from $\eta_{\rm R}\simeq 0.45$ at $n=6$ to $\eta_{\rm R}\simeq 0.6$ at $n=52$ (see Ref.~\cite{BellJCP2020} and related Supplementary Material for further details).}   

At the freezing points of the model systems considered in Fig.~\ref{Fig1} we get roughly $\eta_{\rm R}\sim 5$ and $\lambda_{\rm R}\sim 10$ for LJ and OCP fluids, whereas somewhat higher values are obtained for the HS fluid, $\eta_{\rm R}\sim 7$ and $\lambda_{\rm R}\sim 14$.

\section{Real liquids}

{We have analysed experimental data for the viscosity and thermal conductivity coefficients of seven atomic and molecular liquids along their supercritical isobars provided in the National Institute of Standards and Technology (NIST) Reference Fluid Thermodynamic and Transport Properties Database (REFPROP 10.0)~\cite{Refprop}. In particular, we have selected Ne, Ar, Kr, Xe, and N$_2$ at 10 MPa, CO$_2$ at 30 MPa, and CH$_4$ at 50 MPa. The individual models for viscosity, thermal conductivity, and equation of state for each species can be found in~\cite{NIST}}. 

\begin{figure}
\includegraphics[width=7.5cm]{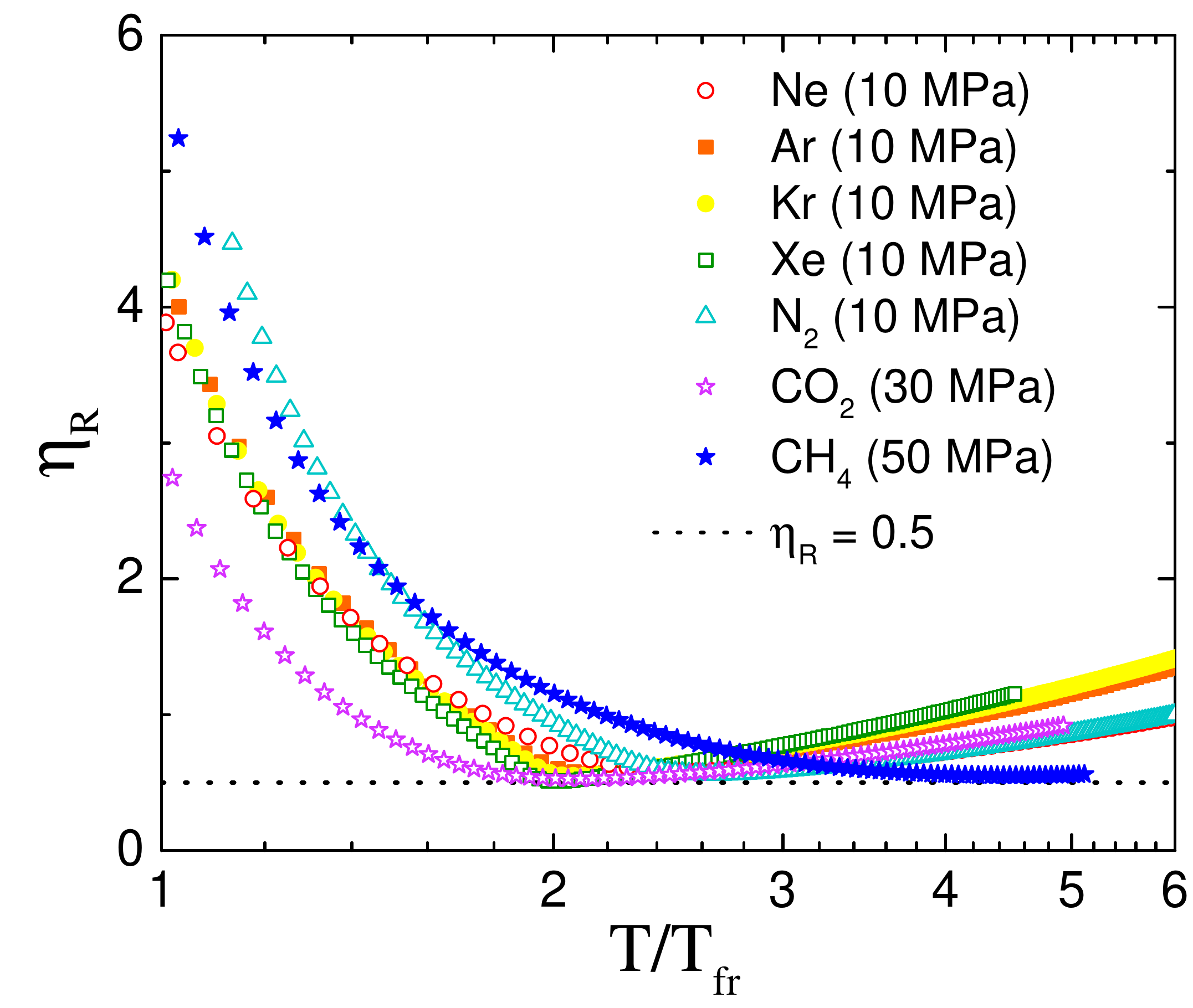}
\caption{(Color online) Reduced shear viscosity coefficient $\eta_{\rm R}$ versus the reduced temperature $T/T_{\rm fr}$ for seven atomic and molecular liquids (Ne, Ar, Kr, Xe, N$_2$, CH$_4$, and CO$_2$) along isobars (see the legend)~\cite{Refprop}. The horizontal dashed line corresponds to $\eta_{\rm R}=0.5$. }
\label{Fig2}
\end{figure}

\begin{figure}
\includegraphics[width=7.5cm]{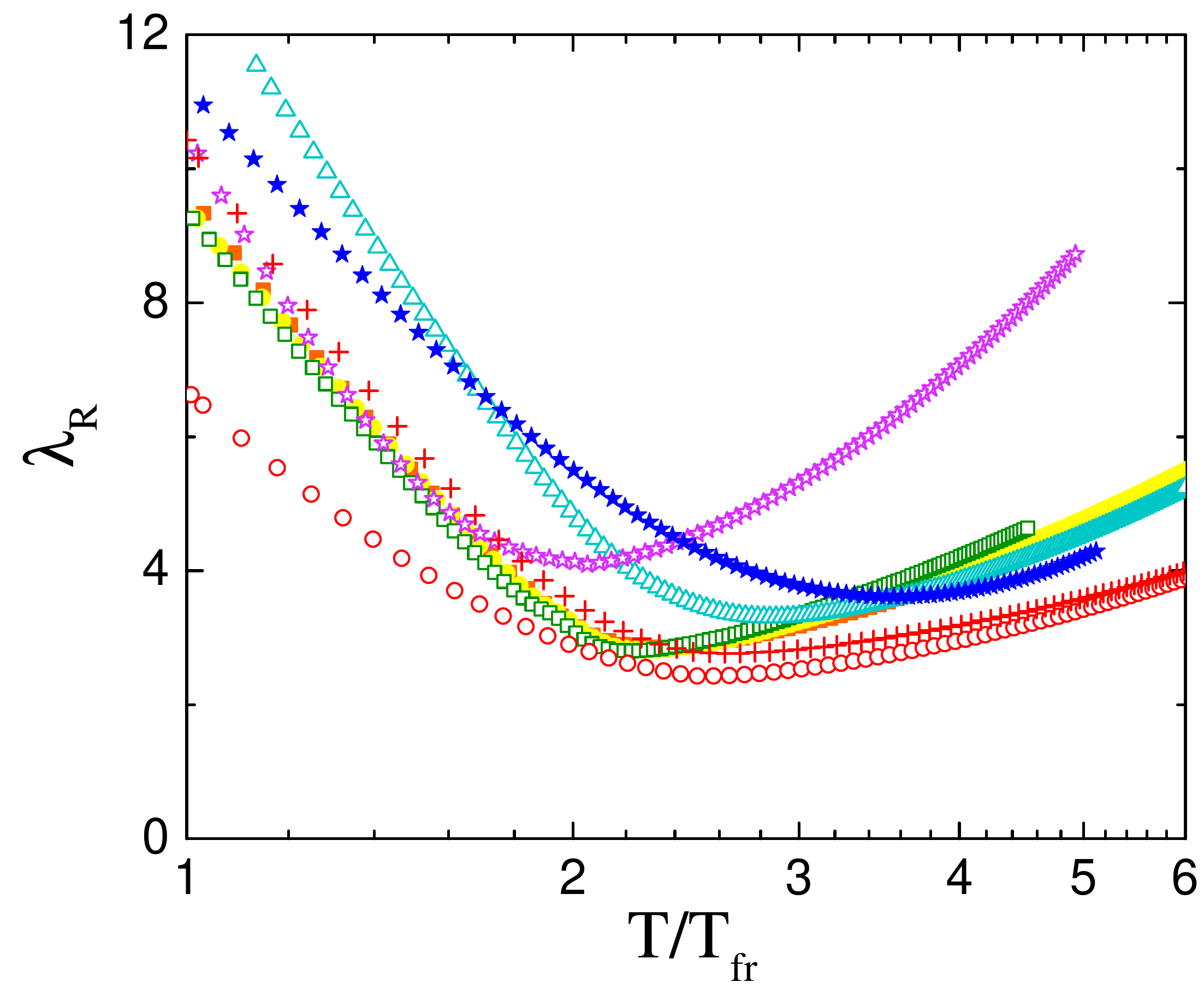}
\caption{(Color online) Reduced thermal conductivity data for $\lambda_{\rm R}$ versus the reduced temperature $T/T_{\rm fr}$ taken from~\cite{Refprop} for the same liquids and pressures as in Fig~\ref{Fig2}. Red crosses correspond to neon liquid data from Ref.~\cite{NIST}.}
\label{Fig3}
\end{figure}

Figure~\ref{Fig2} shows the data for the reduced viscosity, while in Fig.~\ref{Fig3} the reduced thermal conductivity data are plotted. In both cases the horizontal axis corresponds to the temperature normalized to its value at the freezing point, $T/T_{\rm fr}$. Minima are clearly observed (note again that the location of the minima is not universal and can depend on the normalization chosen). {The emerging dependencies are non-universal.  This is of course not surprising, because both transport properties are system-dependent. In addition, the chosen isobars correspond to quite different trajectories on the phase diagrams of considered substances (more similarity could be expected for instance for isobars characterized by the same ratio $P/P_{\rm cr}$, where $P_{\rm cr}$ is the critical pressure). Nevertheless, Figure~\ref{Fig2} demonstrates that the viscosity coefficients of liquified noble gases tend to group around a quasi-universal curve. The low-temperature viscosity of CO$_2$ is below, while low-temperature viscosities of N$_2$ and CH$_4$ are above this curve. Similar picture is observed in Fig.~\ref{Fig3}, except the thermal conductivity data for neon are located considerably lower than those of other liquified noble gases. Puzzled by this behaviour we also plotted the neon thermal conductivity data from an older NIST Chemistry WebBook database~\cite{NIST}. These data are shown by red crosses in Fig.~\ref{Fig3} and are located much closer to those for other noble elements. The discrepancy is significant and probably deserves further attention. Both databases refer to Ref.~\cite{Bewilogua1972} as a main data source for liquid neon thermal conductivity. This point is however beyond the scope of the present paper. For the present consideration it is important that the magnitudes of $\eta_{\rm R}$ and $\lambda_{\rm R}$ at their respective minima do not vary strongly from one substance to another.}

\begin{table}
\caption{\label{Tab1}{Reduced shear viscosity $\eta_{\rm R}^{\rm min}$ and thermal conductivity $\lambda_{\rm R}^{\rm min}$ coefficients at their respective minima. Specific heats $c_{\rm v}/k_{\rm B}$ at the viscosity minima. Reduced shear viscosity and thermal conductivity coefficients at freezing points, $\eta_{\rm R}^{\rm fr}$ and $\lambda_{\rm R}^{\rm fr}$.  Data for seven liquids along indicated isobars are from Ref.~\cite{Refprop}, except the thermal conductivity coefficient of Ne, which is taken from Ref.~\cite{NIST}.}}
\begin{ruledtabular}
\begin{tabular}{crrrrr}
Liquid & $\eta_{\rm R}^{\rm min}$ & $\lambda_{\rm R}^{\rm min}$ & $c_{\rm v}/k_{\rm B}$ & $\eta_{\rm R}^{\rm fr}$ & $\lambda_{\rm R}^{\rm fr}$  \\ \hline
Ne (10 MPa) & 0.58 & 2.77 & 1.72 & 3.97 & 10.4    \\
Ar (10 MPa) & 0.57 &  2.86 & 1.97 & 4.37 & 9.68 \\
Kr (10 MPa) & 0.55 & 2.82  & 2.06 & 4.46 & 9.45 \\
Xe (10 MPa) & 0.51 & 2.80 & 2.18 & 4.34 & 9.37 \\
N$_2$ (10 MPa) & 0.56 & 3.31 & 2.88 & 6.86 & 13.15 \\
CO$_2$ (30 MPa) & 0.53 & 4.09 & 4.73 & 2.92 & 10.51 \\
CH$_4$ (50 MPa) & 0.55 & 3.62 & 4.46 & 5.81 & 11.21 \\
\end{tabular}
\end{ruledtabular}
\end{table}

{The values of $\eta_{\rm R}$ and $\lambda_{\rm R}$ at the minima and at the freezing points are summarized in Table~\ref{Tab1}. The minimal values of $\eta_{\rm R}$ all lie in a narrow range between $0.5$ and $0.6$. The minimal values of shear viscosity are seemingly independent on whether the liquid is atomic or molecular. The minimal values of thermal conductivity do not demonstrate this level of universality. For liquified noble gases $\lambda_{\rm R}^{\rm min}\simeq 2.8$ to a good accuracy, but for molecular liquids it increases considerably. This could be expected, because the thermal conductivity coefficient is proportional to the specific heat, which is higher for molecular systems due to intramolecular degrees of freedom. However, the translational and internal thermal conductivity
contributions cannot be straightforwardly decoupled in a convincing way (see e.g. Ref.~\cite{BellJCED2019} and references therein). In addition, the excess entropy scaling as well as the vibrational model of heat transfer are applicable to monatomic systems. Thus, we cannot quantify this increase in any reasonable way from the theoretical perspective. What experimental results seem to indicate is that the momentum transfer mechanism in the considered regime is much less sensitive to the molecular structure than the mechanism of energy transfer.} 

{The values of $\eta_{\rm R}$ and $\lambda_{\rm R}$ at the freezing points show more scattering. For liquified noble gases $\eta_{\rm R}^{\rm fr} \simeq 4.2\pm 0.3$ and  $\lambda_{\rm R}^{\rm fr}\simeq 9.9 \pm 0.5$ (note that $\eta_{\rm R}\simeq 5-6$ for many liquid metals at their corresponding melting temperatures~\cite{KhrapakAIPAdv2018}). For molecular liquids scattering is more pronounced. In cases considered, $\lambda_{\rm R}$ is 2-3 times larger than $\eta_{\rm R}$ at the freezing point.} Finally, it is observed that a quasi-universal value of the thermal conductivity coefficient at the freezing point ($\sim 10$) is consistent with the estimate we made using a simplified version of the Lindemann melting criterion. It appears that the reduced thermal conductivity at the melting temperature is a more robust quantity than the reduced shear viscosity (which can differ by several times for different substances). 

{In a recent paper the behaviour of he shear viscosity coefficient of normal alkanes n-C$_{N}$H$_{2N+2}$ at the crystal-liquid-vapour triple point has been investigated in detail~\cite{BellJCED2020}.}                    

\section{Discussion and conclusion}
  
The shear viscosity and thermal conductivity coefficients of various model and real fluids usually exhibit minima along trajectories that start from a dilute gaseous state and end in a dense fluid regime near the freezing point. The origin of these minima is the crossover between the gas-like and liquid-like dynamical regimes, which corresponds to different mechanisms of momentum and energy transfer. It has been recently proposed that the kinematic viscosity and thermal diffusivity coefficients have lower bounds, which are fixed by fundamental physical constants~\cite{TrachenkoSciAdv2020,TrachenkoPRB2021}. Moreover, the fundamental lower limits of kinematic viscosity and thermal diffusivity are numerically close~\cite{TrachenkoPRB2021}.

In this paper we have demonstrated that the minima of properly reduced shear viscosity and thermal conductivity coefficients can be estimated using purely classical arguments. This observation is supported by the numerical and experimental data on the transport properties of various model and real fluids {(we should remark here that although classical arguments do apply to real fluids, the details of the interatomic interactions are determined by quantum effects and thus all the absolute measurable quantities such as interatomic separation, melting temperature, Debye frequency, transport coefficients, etc. will in fact depend on quantum constants).}

For monatomic fluids considered (model and real) we obtained $\eta_{\rm R}^{\rm min} = 0.6\pm 0.1$ and $\lambda_{\rm R}\simeq 2.8 \pm 0.2$. The only exception is the one-component plasma fluid, where these minimal values are considerably lower ($\eta_{\rm R}\simeq 0.3$ and $\lambda_{\rm R}\simeq 1.2$). This is related to the extremely soft and long-ranged character of the Coulomb potential, in agreement {with the dependence of the transport coefficients minima on the interaction potential softness reported previously. From the presented evidence we can hypothesise that the OCP values can serve as classical lower bounds on the viscosity and thermal conductivity coefficients of liquids. In non-reduced units we get:
\begin{equation}
\eta\gtrsim 0.3\frac{\sqrt{k_{\rm B}Tm}}{n^{2/3}}, \quad\quad \lambda \gtrsim 1.2 \frac{\sqrt{k_{\rm B}T}}{\sqrt{m}n^{2/3}}.
\end{equation}
}  

For molecular liquids, the minimal value of $\eta_{\rm R}$ is close to that in monatomic liquids. The minimal value of $\lambda_{\rm R}$ somewhat increases, presumably due to the contribution from additional degrees of freedom to the energy transfer. 

We have also observed that the values of $\eta_{\rm R}$ and $\lambda_{\rm R}$ at the respective freezing points of the considered substances are less universal than those at the minima. At the same time, the vibrational model of heat transfer combined with the simplest version of the Lindemann melting criterion allows for a rough but reasonable estimate of the reduced heat conductivity coefficient near the fluid-solid phase transition. 

Altogether, these findings can represent an important step to a better understanding of transport properties of liquids across their phase diagrams.         

Data sharing is not applicable to this article as no new data were created or analyzed in this study. 




\bibliography{SE_Ref}

\end{document}